Accepted Manuscript. Published version available in International Journal of Human-Computer Interaction. DOI: 10.1080/10447318.2025.2474469.

# Teachers' Vocal Expressions and Student Engagement in Asynchronous Video Learning

Hung-Yue Suen[1]; Yu-Sheng Su[2,3,4]*

[1] Department of Technology Application and Human Resource Development, National Taiwan Normal University, Taipei City, Taiwan

[2] Department of Computer Science and Information Engineering, National Chung Cheng University, Chiayi, Taiwan

[3]Advanced Institute of Manufacturing with High-tech Innovations, National Chung Cheng University, Taiwan

[4] Department of Computer Science and Engineering, National Taiwan Ocean University, Taiwan

*Corresponding author: Yu-Sheng Su

E-mail address: ccucssu@gmail.com

## Abstract

Asynchronous video learning, including massive open online courses (MOOCs), offers flexibility but often lacks students' affective engagement. This study examines how teachers' verbal and nonverbal vocal emotive expressions influence students' self-reported affective engagement. Using computational acoustic and sentiment analysis, valence and arousal scores were extracted from teachers' verbal vocal expressions, and nonverbal vocal emotions were classified into six categories: anger, fear, happiness, neutral, sadness, and surprise. Data from 210 video lectures across four MOOC platforms and feedback from 738 students collected after class were analyzed. Results revealed that teachers' verbal emotive expressions, even with positive valence and high arousal, did not significantly impact engagement. Conversely, vocal expressions with positive valence and high arousal (e.g., happiness, surprise) enhanced engagement, while negative high-arousal emotions (e.g., anger) reduced it. These findings offer practical insights for instructional video creators, teachers, and influencers to foster emotional engagement in asynchronous video learning.

## 1. INTRODUCTION

Asynchronous video learning, a method of learning content via prerecorded videos, has emerged as a cornerstone of modern education, particularly in the context of distance learning and massive open online courses (MOOCs). While this mode of learning offers unparalleled flexibility and accessibility, it presents significant challenges in fostering student engagement, particularly affective engagement, due to the lack of real-time interaction between teachers and students. This study aims to explore how teachers' verbal and nonverbal vocal emotive expressions can establish a social presence and, in turn, enhance students' affective engagement in MOOCs.

Asynchronous video learning empowers students with flexibility and control over their learning journey, driving its widespread adoption in corporate training, adult education, and higher education (Zeng & Luo, 2023). Reflecting this paradigm shift, the MOOC market is projected to grow from USD 14.75 billion in 2022 to USD 149.42 billion by 2029 (Maximize Market Research, 2023). Despite its advantages, this learning mode lacks interactivity between teachers and students, posing challenges in educational practice and research (Alemayehu & Chen, 2023; Garcia & Yousef, 2023). Teachers in asynchronous settings struggle to monitor and respond to students' reactions in real time, making it difficult to adjust communication strategies accordingly (Mershad & Said, 2022). This limitation highlights the pressing need for research on strategies to effectively engage students in asynchronous video learning, particularly through prerecorded video lectures (Ma et al., 2023).

Student engagement, a critical element in asynchronous video learning, is closely linked to persistent learning interest and learning outcomes (Lin et al., 2019; Zhao & Khan, 2022). Among the three dimensions of learning engagement delineated by Walker and Koralesky (2021)—affective, behavioral, and cognitive-affective engagement poses unique challenges in asynchronous settings due to the limited teacher-student interactions.

The “affective” dimension captures students' emotional responses, significantly influencing their motivation. The "behavioral" side includes things like participating in discussions and doing assignments, and the "cognitive" side is about how much students have to think to complete tasks like analyzing information and putting their knowledge to use. However, affective engagement is a bit different, especially in asynchronous learning. Because there's less direct contact between teachers and students compared to in-person classes or live online sessions (Daher et al., 2021; Garcia & Yousef, 2023), figuring out how to promote affective engagement in these environments is critical and something that needs more study (Walker & Koralesky, 2021).

Building on the Community of Inquiry (CoI) framework (Borup et al., 2012), three core elements—social presence, cognitive presence, and teaching presence—are essential for a successful learning experience in online environments. Social presence in virtual settings is about participants recognizing each other as real people and forming connections. Cognitive presence is the extent to which learners build and confirm understanding through critical reflection. Teaching presence involves designing, facilitating, and guiding learning activities to achieve meaningful outcomes.

The importance of social presence is magnified in asynchronous learning due to the inherent limitations of teacher-student interaction. Research shows that students engage more deeply when they perceive their teacher as a "real" presence, conveyed through emotional expressions like facial cues and vocal tones (Chew, 2022). The link between a teacher's emotional expression and student engagement is mediated by social presence, according to Richardson and Lowenthal (2017). They suggest that appropriate emotional cues, whether in real-time or asynchronous communication, foster immediacy and intimacy, which, in turn, strengthens students' affective engagement. In asynchronous video learning, students focus primarily on the learning materials and instructions rather than the teachers themselves, making teachers' verbal and nonverbal vocal expressions essential for fostering affective engagement (Wang et al., 2022). Verbal vocal expressions are the words spoken by teachers, while nonverbal vocal expressions include tone, pitch, and intonation. Both help convey emotions and engage students in live and asynchronous learning (Wang et al., 2022).

Past studies indicate that teachers' emotional expressions significantly influence student affective engagement (Dixson, 2010; Wang, 2022). This finding is supported by the emotional response theory (Mottet et al., 2006), which posits that teachers' emotional expressions, observable in their demeanor, are detectable by students and thus trigger corresponding emotional responses among students (Liu et al., 2019). This observation resonates with the concept of emotion contagion, which posits that emotions can spread automatically between people, prompting similar reactions (Hatfield et al., 1993). These reactions are characterized by two key dimensions: "valence" (the pleasantness or unpleasantness of the feeling) and "arousal" (ranging from alertness and responsiveness to a state of calmness) (Mottet et al., 2006).

Research has shown that speakers' verbal and nonverbal vocal expressions with positive and greater arousal emotions induce the greatest emotion contagion effect among all vocal expressions (Russell, 2003). When applied to education, teachers' expressions often predict specific student behaviors more accurately than students' emotional responses, thus influencing attention, memory, motivation (Fanselow, 2018), satisfaction, and performance (Iskrenovic-Momcilovic, 2018).

A critical question remains unanswered: Do the insights into teacher-student interactions extend to the rapidly growing field of asynchronous video learning? Perry and colleagues (2024) highlight this issue, calling for more research into how instructors are presented in videos and how this affects student reactions and learning—an area that remains largely unexplored. This gap underscores the urgent need to investigate human-computer interaction in educational contexts further. Recent studies have echoed this call, emphasizing the importance of addressing these unanswered questions (Chiu, 2022; Zeng & Luo, 2023). Responding to this need, the current study explores the dynamics of teachers' verbal and nonverbal vocal emotive expressions in prerecorded video instruction, aiming to bridge this gap in the literature.

As shown in Table 1, it explored verbal expressions characterized by valence (positive or negative tone) and arousal (energy level) and six nonverbal vocal expressions: happiness and surprise (positive, high activation), anger (negative, high activation), fear and sadness (negative, moderate activation), and neutral (neutral, moderate activation). Affective engagement, the dependent variable, was assessed through students' self-reported emotional connection to and involvement with the instructional video.

Table 1. Variables and Measurement Tools

| Variables | Measures | Tools |
|---|---|---|
| Independent variables | | |
| Verbal Emotive Expressions (e.g. spoken words) | Valence Scores (1-9)<br>Arousal Scores (1-9)<br>Valence*Arousal Scores (1-81) | Sentiment Analysis |
| Vocal Emotive Expressions (e.g. tone, pitch, and intonation) | Anger probability (0-100%)<br>Fear probability (0-100%)<br>Happiness probability (0-100%)<br>Neutral probability (0-100%)<br>Sadness probability (0-100%)<br>Suprise probability (0-100%) | Acoustic Analysis |
| Dependent variables | Affective Engagement (1-5) | Self-reported Survey |

To this end, the study seeks to answer the following research questions:

1. Do teachers' verbal vocal expressions with high positive valence and arousal increase students' affective engagement?

2. Do teachers' nonverbal vocal expressions with high positive valence and arousal increase students' affective engagement?

3. Do teachers' nonverbal vocal expressions with high negative valence and arousal decrease students' affective engagement?

4. Which type of vocal expression impacts students' affective engagement more—verbal or nonverbal?

By addressing these questions, this study aims to enhance our understanding of effective teachers' verbal and vocal emotive expressions in asynchronous video learning environments, including MOOCs. The findings are expected to provide actionable insights for designing asynchronous video-based instruction and inform best practices in this rapidly evolving domain of human-computer interaction.

## 2. LITERATURE REVIEW

To understand how verbal and vocal emotive expressions can establish a social presence and, in turn, enhance learners' affective engagement, we build on the abovementioned four research questions and extensively review the relevant literature, focusing on the role of teachers' vocal emotive expressions in asynchronous video learning environments.

## 2.1 Teachers' verbal vocal emotive expression and students' affective engagement

Five experiments conducted by Kraus (2017) demonstrated that vocal cues provide a greater degree of accuracy in perceiving others' emotions compared to facial expressions, attributing this finding to cognitive limitations. Furthermore, individuals can discern emotions from both verbal content (what is said) and nonverbal vocal cues (how it is said), independent of facial expressions (Kraus, 2017; Simon-Thomas et al., 2009).

Speech is the most rapid and natural medium through which to convey verbal content and nonverbal cues in human communication (El Ayadi et al., 2011). Spoken words in speech convey not only the cognitive messages and behavioral intentions of the speaker but also various emotional states, inducing listeners' emotional valence and arousal (Beukeboom & Semin, 2006; Liebenthal et al., 2016).

Affective events theory suggests that emotions play a central role in individuals' reactions to experiences, with verbal expressions acting as significant affective events that influence students' emotional responses and, in turn, their engagement levels (Weiss & Cropanzano, 1996). Moreover, linguistic theory highlights the influence of language in shaping emotional responses. Accordingly, teachers' choice of words can greatly impact students' perceptions and emotional engagement, thus creating a learning environment that is either supportive or detrimental (Immordino-Yang & Damasio, 2007)

Russell's (1980) circumplex model of affect offers a useful framework for understanding emotional perception, categorizing it along two key dimensions: valence and arousal. Valence distinguishes emotions based on their pleasantness (or unpleasantness), while arousal captures the spectrum from calm to excited states. Russell (2003) states this two-dimensional model has shown remarkable adaptability across various cultural settings, highlighting its broad applicability. It indicates that vocal emotions characterized by positive valence and high arousal often leave a stronger emotional impact.

This idea finds support in communication research. Listeners demonstrate an attentional reflex, known as an orienting response, toward positive words, a reaction that is typically diminished or absent for neutral or negative words (Lee & Potter, 2020). Furthermore, positive words are not only more likely to capture attention but are also encoded more effectively. Lee and Potter (2020) observed that positive words are processed more efficiently than neutral words, with negative words being the least effectively encoded.

This phenomenon is not limited to English-language contexts (Stevenson et al., 2007). Ho et al. (2015) examined 160 Chinese words and observed that these words elicited different levels of emotional valence among adolescents in Hong Kong and Mainland China, illustrating the subtle cultural differences in emotional expression. In Taiwan, Lee et al. (2022) introduced the Chinese EmoBank, a significant tool for researchers studying emotional language. Their work involved a comprehensive analysis combining semantic insights with affective computing, evaluating a vast collection of Chinese linguistic units, each rated for valence and arousal.

According to Cavanagh et al. (2014), students' assessments of preservice teachers' presentations were significantly affected by the teachers' word choices in recorded videos. Based on these findings, we propose that teachers' spoken words characterized by more positive valence and higher arousal in asynchronous video learning can enhance students’ affective engagement. Therefore, we suggest the following hypothesis:

*H1: The frequency of teachers' verbal vocal expressions with positive valence and high arousal positively influences students' affective engagement.*

## 2.2 Teachers' nonverbal vocal emotive expression and student affective engagement

Speakers convey a complex array of emotions that extends beyond the literal meaning of their words. This emotional content is expressed through various nonverbal vocal cues, including changes in tone, volume, speech pace, and timbre (Juslin & Laukka, 2003; Laukka, 2017). These nuanced vocal variations elicit distinct emotional responses in listeners, encompassing a spectrum of emotions such as happiness, surprise, anger, fear, sadness, and neutrality. Specifically, happiness and surprise are associated with positive emotional states and heightened physiological arousal, whereas anger corresponds to negative emotions but is characterized by a similarly elevated level of arousal (Gunes et al., 2011).

Several theoretical perspectives help explain how a teacher's nonverbal vocal cues shape student affective engagement in educational contexts. For instance, paralanguage theory highlights the powerful role of nonverbal speech characteristics, such as tone, pitch, and rhythm, in shaping listener perceptions and emotions (Trager, 1958). This perspective suggests that subtle vocal variations by a teacher can foster a more dynamic and engaging learning atmosphere. Similarly, self-determination theory posits that an emotionally warm and empathetic vocal tone can strengthen the connection between teachers and students, thereby enhancing students' engagement and motivation to learn (Ryan & Deci, 2000).

Furthermore, the theory of voice and emotion posits that nonverbal vocal cues convey emotional states, affecting listeners' feelings (Juslin & Laukka, 2003). Thus, when emotionally attuned, a teacher's nonverbal vocal cues can create a conducive emotional environment for learning, promoting student engagement.

Chew (2022), in a qualitative analysis, highlighted the importance of teachers' nonverbal vocal expressions as essential pedagogical tools for engaging students in real-time online classes. Moreover, Yuan et al. (2021) provided empirical evidence that these expressions can benefit students' social presence and enjoyment and alleviate boredom in asynchronous learning contexts. Russell (2003) observed that the interplay of valence and arousal in nonverbal vocal expressions significantly influences their role in emotional contagion. As Tursunov et al. (2019) explain, vocal emotions can be positioned along a valence spectrum, ranging from positive emotions like happiness and surprise to negative ones like fear, sadness, and anger, with neutrality at the midpoint. These emotions also differ in their level of arousal, with happiness, anger, and surprise representing the high end of the spectrum, followed by fear, neutral, and then sadness at the low end. It is worth noting that happiness, anger, and surprise exhibit the most distinct valence and the highest levels of arousal.

For example, vocal expressions of happiness and surprise, with their positive valence and increased arousal, correlate with heightened affective engagement. In contrast, vocal expressions of anger, marked by negative valence and high arousal, can lead to decreased affective engagement. Therefore, asynchronous video learning materials featuring teachers expressing positive valence with greater arousal through nonverbal cues can enhance emotional connectivity. Conversely, the perception of negative valence with higher arousal in a teacher's tone can hinder such engagement. On the basis of these insights, the subsequent hypotheses are suggested to guide future investigations in asynchronous video learning:

*H2: The frequency of teachers' nonverbal vocal expressions with positive valence and high arousal positively influences students' affective engagement.*

*H3: The frequency of teachers' nonverbal vocal expressions with negative valence and high arousal negatively influences students' affective engagement.*

### 2.3 Effects of Teachers' verbal vs. nonverbal vocal expressions on student affective engagement

Dual coding theory (Paivio, 1971) suggests that emotional expressions are processed via the following two distinct pathways: a cognitive route for verbal content and an automatic, affective route for nonverbal cues. This theory posits that nonverbal vocal cues are processed more directly than verbal cues, with the former thus having an immediate impact on affective engagement. Nonverbal vocal cues, which require less cognitive interpretation, convey emotions more efficiently and intuitively than do verbal vocal cues. These cues are processed by the brain's limbic system, are responsible for emotional responses, and thus have a more direct and profound effect on listeners than does the cognitive processing involved with verbal cues (Horan et al., 2012; Scherer, 1986).

Mehrabian's (1971) communication rule indicates that the impact of a message comes 7% from words, 38% from nonverbal vocal elements, and 55% from other nonverbal elements, emphasizing the importance of nonverbal communication in face-to-face interactions. However, in the context of one-way communication, which is typical of video instruction—where teachers may not be visible, and students may not constantly watch the screen—the emotional impact of nonverbal vocal elements likely exceeds that suggested by Mehrabian's rule, overshadowing verbal elements. Hsu and colleagues (2021) demonstrated that a speaker's nonverbal vocal cues have a significant emotional impact, offering a direct and universally understandable channel for emotional communication that surpasses the expressive limitations of verbal language, thus enhancing audience affective engagement with video content. The immediacy of nonverbal cue processing highlights the critical role of such cues in defining a video's emotional tone, rendering these cues essential for speakers aiming to communicate deeper emotional states.

In asynchronous video learning, where learner–instructor interactions are limited to fewer communication channels, learners may rely predominantly on the instructor's vocal cues and presented materials (Fabriz et al., 2021). As such, a teacher's nonverbal vocal cues in asynchronous online learning contexts may significantly influence students' affective engagement more than do spoken words. On the basis of these insights, we hypothesize the following in asynchronous video learning:

*H4: Teachers' nonverbal vocal expressions have a greater influence on students' affective engagement than do their verbal vocal expressions.*

## 3. METHODS

This study examined the impact of teachers' verbal and nonverbal vocal emotive expressions in asynchronous video learning content on adult students' affective engagement. In collaboration with four prominent MOOC platforms through industry-academic partnerships, we collected prerecorded course videos, all de-identified by the platforms, to protect the instructors' privacy. The average video length was approximately 1 hour and 10 minutes. To analyze teachers' verbal vocal expressions, we applied sentiment analysis using established academic datasets to calculate valence and arousal scores based on the word choices in the videos. For nonverbal vocal expressions, we employed acoustic analysis combined with a deep learning (DL) model trained on existing datasets to automatically detect the frequency of various emotions, such as happiness and anger, in the teachers' speech. Additionally, the MOOC platforms used a five-item survey, tested for reliability and accuracy, to measure students' affective engagement right after they watched each video.

With the assistance of these platforms, videos were anonymized using specialized software. Via DL techniques, we evaluated the visibility of teachers on screen and analyzed the verbal and nonverbal vocal emotive expressions conveyed through teachers' vocal cues. Teachers' gender, course subject, and course length were also automatically extracted by the platform systems. Following the completion of each video, an online survey was administered on the platforms to measure students' affective engagement as part of the final post-class feedback. This study used multiple regression analysis to test the research hypotheses. The research design is illustrated in Figure 1.

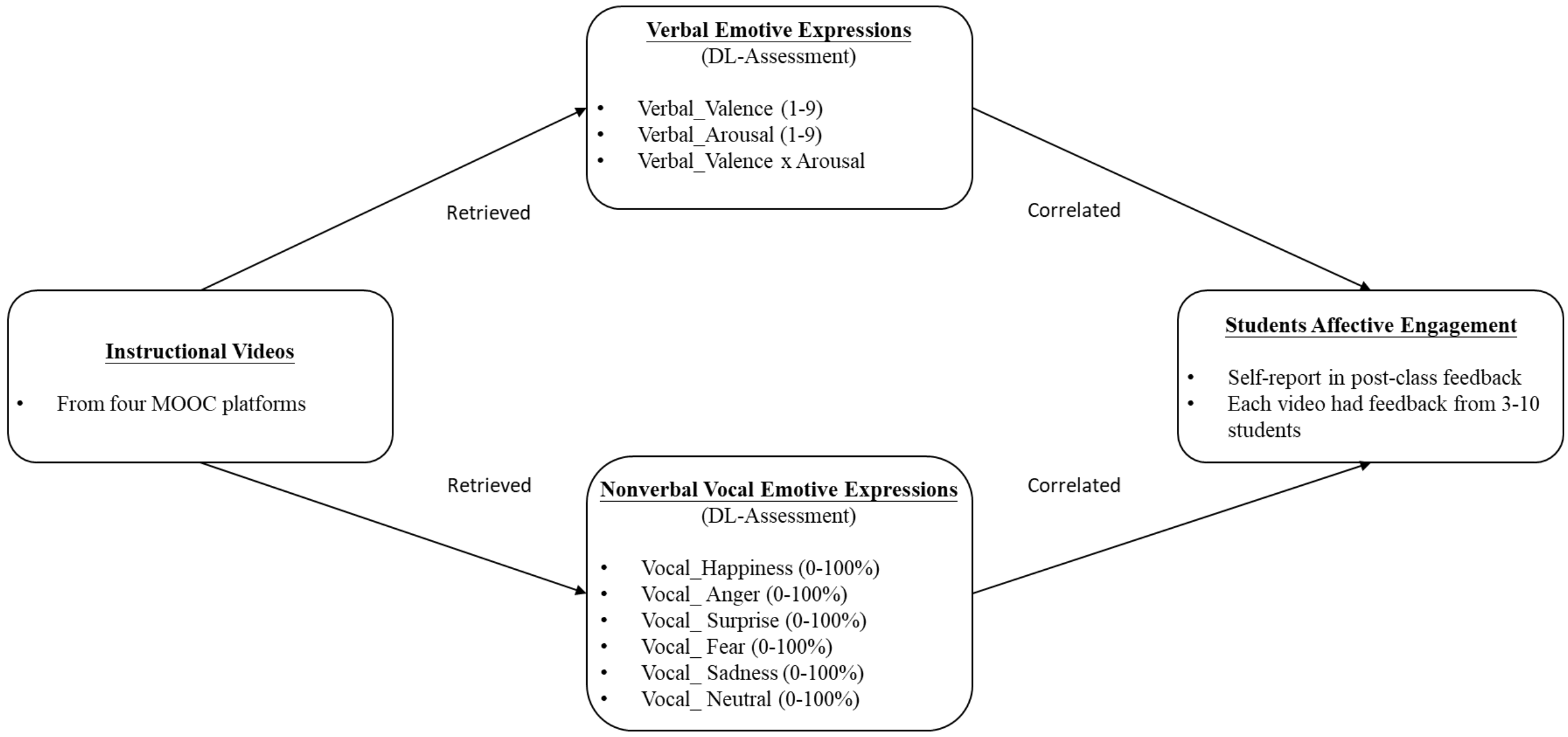

Figure 1. Research Design

### 3.1 Data collection

Four Taiwan-based MOOC platforms—Hahow, Kvalley, TibaMe, and TKB—offer a variety of courses for adult learners and those seeking higher education using primarily prerecorded videos. Courses were categorized into sciences and social sciences, with video quality above 1280×720 resolution. In some videos, teachers' faces appeared for up to one-quarter of the video's length, whereas other videos did not show the instructors' faces at all, with varying durations.

In this study, 210 valid video samples from unique teachers, defined as videos accurately transcribed and sentiment-analyzed using DL, were collected and analyzed. The sample included 152 male (72%) and 58 female (28%) teachers. Most courses (179, 85%) were in the social sciences, with the rest (31, 15%) being in science. Lecture durations ranged from 10 to 698 minutes, averaging 131 minutes. Teachers appeared on screen in 103 videos (49%) and were absent in 107 (51%).

To evaluate students' affective engagement, the platforms invited students who had completed each video lecture to participate in a post-class survey. Each video was assessed by at least three different students, ensuring diverse perspectives. In total, 738 unique students participated in the surveys, providing an average of 3.51 evaluations per video.

Participants were aged 18 to 45 years, mainly in the 26-30 (33%) and 31-35 (27%) year brackets. Female respondents made up 58% of the survey population, whereas male respondents accounted for 42%. Most respondents held bachelor's degrees (65%), with the remaining 35% holding master's degrees.

### 3.2 Data measures

This research employed regression analysis to assess the proposed hypotheses. The independent variables included emotive Verbal_Valence, Verbal_Arousal, and Verbal_Aalence*Arousal from spoken words to represent vocal emotive contagion scores, along with the following six distinct nonverbal vocal expressions: happiness, anger, surprise, fear, sadness, and neutral. The dependent variable was students' self-reported affective engagement, which was averaged for each video.

### *3.2.1 Teachers' verbal vocal emotive expression*

An automatic speech recognition (ASR) system, "Whisper," developed by OpenAI (2022) and integrated with natural language processing (NLP), was used to convert spoken language from video feeds into Chinese text. Empirical tests indicated that in addition to minor issues with traditional-simplified Chinese conversions (e.g., the pronunciation of "zh"), its error rate was lower than that of other Chinese ASR solutions (Heikinheimo, 2023).

After transcribing each video, we used Jieba-tw (2023) for word segmentation and matched these words with the emotion-inducing word database from Chinese EmoBank (Lee et al., 2022) using Python. This approach allowed us to extract emotional Verbal_Valence (0.92 to 9.00) and Verbal_Arousal (1.00 to 9.00) scores for each word via sentiment analysis. We then multiplied the valence and arousal scores to derive the emotional contagion score for each word. These scores were summed and divided by the total number of words analyzed from the video to quantify the intensity of emotional engagement elicited by teachers' spoken words.

As an example, imagine a video where two words influenced emotional contagion. One word had a valence rating of 2 and an arousal rating of 3, resulting in a score of 6 (the product of valence and arousal). Another word in the same video had a valence of 5 and an arousal of 4, producing a score of 20. The video's overall emotional contagion score was then calculated by summing these individual scores (26) and dividing by the number of contributing words (2), resulting in a final score of 13.

### *3.2.2 Teachers' nonverbal vocal emotive expression*

To analyze teachers' vocal expressions of emotion, we employed a deep learning model combining convolutional neural networks (CNNs) and long short-term memory (LSTM) networks, drawing on the approach outlined by Ye et al. (2022). This model was trained using the CASIA dataset from the Chinese Academy of Sciences, Institute of Automation (Bao et al., 2014). Our approach to emotion detection involved extracting a comprehensive set of 193 acoustic features, including Chroma features, mel-scaled spectrograms, mel-frequency cepstral coefficients (MFCCs), spectral contrast, and Tonnetz representations (Sharanyaa et al., 2023; Su et al., 2020; Su et al., 2021a; Su et al., 2021b; Su et al., 2022), capturing frequencies from 20 Hz to 20 kHz. To enhance the signal-to-noise ratio, audio segments were processed in five-second chunks with a four-second overlap, and segments exhibiting confidence levels below the 0.7 threshold were omitted (Allison et al., 2022).

The CASIA, featuring vocal recordings from Chinese actors in various emotional states, was allocated into training and testing portions with an 80/20 distribution. This dataset includes six primary emotions—Vocal_Happiness, Vocal_Anger, Vocal_Surprise, Vocal_Fear, Vocal_Sadness, and Vocal_Neutral—rated on a 0 to 100% intensity scale. Our testing results achieved an accuracy (ACC) rate of 81%, approaching the 84% ACC cited by Ye et al. (2022) for video-based emotion recognition using the CASIA dataset.

#### *3.2.3 Student affective engagement*

We employed Dixson’s (2010) emotional engagement scale to assess affective engagement. Participants rated the following items on a 5-point Likert scale, with endpoints of "Not at all characteristic of me" (1) and "Very characteristic of me" (5):

“I put forth effort while engaging with the video.”

“I found ways to make the materials relevant to me.”

“I applied the materials to my life.”

“I found ways to make the materials interesting.”

“I really desired to learn the content.”

This scale was selected for its high reliability and strong relevance to online learning contexts, as noted by Henrie et al. (2015). It has demonstrated high reliability, with Cronbach’s alpha ranging from 0.86 to 0.95 across different sample tests, and has been validated through its significant correlation with applied learning behaviors ($r = .48$, $p < .01$) (Dixson, 2015).

Its effectiveness and broad applicability across diverse online learning environments have been demonstrated by numerous studies, further supporting its use for measuring engagement in MOOCs and other forms of asynchronous video instruction (e.g., Farrell & Brunton, 2020; Vo & Ho, 2024).

### 3.3 Data analysis

We analyzed the relationship between teacher vocal cues and student affective engagement using a two-part approach. First, we conducted correlation analysis to determine which variables to include in our subsequent regression models. Then, we used hierarchical linear regression, entering Verbal_Valence and Verbal_Arousal in the first block, the interaction term (Verbal_Valence*Arousal) in the second, and the frequencies of six vocal expressions (Vocal_Happiness, Vocal_Anger, Vocal_Surprise, Vocal_Fear, Vocal_Sadness, Vocal_Neutral) in the final block. Student affective engagement was the dependent variable.

## 4. RESULTS

### 4.1 Reliability analysis

To gauge student affective engagement, we used post-class surveys administered immediately after each video lecture within a MOOC unit. These surveys typically included assessments of content mastery, such as quizzes or tests, alongside questionnaires designed to gather feedback on the overall learning experience. For this particular study, we incorporated a five-item questionnaire specifically targeting affective engagement into this standard post-class survey. This scale demonstrated strong internal consistency, with a Cronbach's alpha (α) of 0.83. With multiple students evaluating each teacher, the average intraclass correlation coefficient (ICC) was 0.66, indicating substantial consistency and reliability.

Independent variables were derived from biographical data and assessed through DL models, which achieved an ACC of 81% in recognizing perceived vocal emotions on the basis of the CASIA dataset, thus satisfying the benchmark ACC of 70–80% (Schuller et al., 2018).

## 4.2 Correlation analysis

Research indicates that the on-screen visibility of teachers affects student engagement in online classes (Chew, 2022; Henderson & Schroeder, 2021). Additionally, teacher gender may influence students' emotional perceptions (Leung et al., 2018). Course subject and length also impact student affective engagement, as variations in these factors can affect engagement levels (Akiha et al., 2018). Therefore, this study tested whether the above four variables influence student affective engagement. If significant correlations were found, then these variables were included as control variables in the regression model to test our hypotheses.

To understand the relationships among all the variables, we first conducted a correlation analysis, as shown in Table 2. The correlation matrix revealed that teacher visibility, course subject, video length, and teacher gender did not affect students' affective engagement. Consequently, these variables were not included in the hypothesis testing because of their lack of a significant relationship with the dependent variable.

Table 2. Correlation Analysis

| Variable | Mean | SD | 1 | 2 | 3 | 4 | 5 | 6 | 7 | 8 | 9 | 10 | 11 | 12 | 13 |
|---|---|---|---|---|---|---|---|---|---|---|---|---|---|---|---|
| 1. Visibility[a] | 1.49 | 0.501 | — | | | | | | | | | | | | |
| 2. Gender[b] | 1.724 | 0.448 | -.012 | — | | | | | | | | | | | |
| 3. Subject[c] | 1.852 | 0.356 | .167* | -.197** | — | | | | | | | | | | |
| 4. Length | 130.633 | 100.152 | -.186** | .102 | -.021 | — | | | | | | | | | |
| 5. Verbal_Valence | 5.53 | 0.345 | -.096 | -.077 | -.130 | -.014 | — | | | | | | | | |
| 6. Verba_Arousal | 4.469 | 0.231 | .051 | -.125 | .136* | -.154* | .119 | — | | | | | | | |
| 7. Verbal_Valence*Arousal | 24.729 | 2.224 | -.039 | -.135 | -.012 | -.098 | .818*** | .665*** | — | | | | | | |
| 8. Vocal_Happiness | 0.159 | 0.053 | -.088 | .081 | .055 | .046 | .036 | -.034 | .016 | — | | | | | |
| 9. Vocal_Anger | 0.072 | 0.028 | -.091 | .042 | -.031 | -.011 | .054 | -.053 | .009 | -.041 | — | | | | |
| 10. Vocal_Surprise | 0.2 | 0.096 | .007 | -.025 | .106 | .013 | .120 | .092 | .151* | .148* | -.194* | — | | | |
| 11. Vocal_Fear | 0.138 | 0.074 | .020 | .006 | -.004 | -.016 | -.042 | -.012 | -.034 | -.128 | .057 | -.039 | — | | |
| 12. Vocal_Sadness | 0.131 | 0.085 | .078 | -.018 | -.084 | -.035 | -.009 | .021 | .011 | -.075 | -.039 | .075 | .097 | — | |
| 13. Vocal_Neutral | 0.63 | 0.142 | -.038 | .018 | -.118 | .089 | -.043 | -.027 | -.043 | -.063 | -.049 | -.008 | .007 | .167* | — |
| 14. Affective Engagement | 4.082 | 0.361 | -.052 | -.106 | .058 | .074 | .080 | .034 | .083 | .328*** | -.332*** | .377*** | -.045 | .026 | -.062 |

*p<.05, **p<.01, and ***p<.001.

[a]1 =Off-screen, 2=On-screen.

[b]1 =Female, 2=Male.

[c]1 =Science, 2=Social Sciences.

Correlation analysis indicated that teachers expressing vocal emotions of happiness and surprise were perceived by students as eliciting greater affective engagement, whereas those displaying anger had the opposite effect. The calculation of verbal contagion by multiplying valence and arousal implied a high degree of correlation between these levels. Additionally, there was no significant correlation between the valence and arousal of spoken words or among different vocal emotions, except for vocal surprise, where a very low-level correlation was noted, indicating the distinctiveness of these variables.

## 4.3 Linear regression analysis

To test the hypotheses, we used hierarchical linear regression. Initially, we introduced Verbal_Valence and Verbal_Arousal as independent variables in the first layer (Model A) and added the interaction term Verbal_Valence*Arousal in the second layer (Model B). In the last layer (Model C), we added six types of nonverbal vocal emotive expressions to examine changes in explained variance.

As detailed in Table 3, the analysis revealed that Model A, which focuses on the valence and arousal of teachers' spoken words, did not significantly contribute to affective engagement ($p$ =.271 and 720), explaining only 0.3% of the variance according to the adjusted $R^2$ value. Adding the interaction of valence and arousal (Verbal_Valence*Arousal) in Model B increased the adjusted $R^2$ value by approximately 0.2%, with higher standardized coefficients for verbal valence and arousal. The coefficients of Verbal_Valence and Verbal_Arousal changed from positive to negative in Model B because of the altered interpretation of these main effects in the presence of the interaction. However, this interaction did not have a significant effect on affective engagement; thus, H1 was not supported.

Conversely, incorporating six nonverbal vocal emotive expressions in Model C showed that teachers' nonverbal cues, particularly those with positive valence and higher arousal, such as happiness and surprise, significantly increased student affective engagement. A negative valence with higher arousal (anger) negatively impacted engagement. Nonverbal vocal expressions with less distinct valence and arousal, such as fear, sadness, and neutral expressions, did not significantly affect engagement, thus supporting H2 and H3.

Comparing Models B and C, including nonverbal vocal expressions augmented the explanatory power for affective engagement by 25.6% over verbal emotive expressions alone, according to the adjusted $R^2$ value, thus affirming H4.

Table 3. Hierarchical Regression Analysis

**Dependent Variable: Affective Engagement**

| **Model** | **Variable** | **Standardized Coefficient** | **t** | **p** | **95% CI** | | **$R^2$** | **Adjusted $R^2$** |
|---|---|---|---|---|---|---|---|---|
| | | | | | Lower | Upper | | |
| A | Verbal_Valence | 0.077 | 1.104 | 0.271 | -0.063 | 0.225 | 0.007 | 0.003 |
| | Verbal_Arousal | 0.025 | 0.359 | 0.720 | -0.176 | 0.254 | | |
| B | Verbal_Valence | -0.429 | -0.558 | 0.577 | -2.038 | 1.139 | 0.009 | 0.005 |
| | Verbal_Arousal | -0.364 | -0.614 | 0.540 | -2.391 | 1.255 | | |
| | Verbal_Valence*Arousal | 0.675 | 0.661 | 0.509 | -0.217 | 0.437 | | |
| C | Verbal_Valence | 0.331 | 0.492 | 0.623 | -1.045 | 1.739 | **0.293** | **0.261** |
| | Verbal_Arousal | 0.211 | 0.407 | 0.684 | -1.269 | 1.929 | | |
| | Verbal_Valence*Arousal | -0.375 | -0.418 | 0.676 | -0.348 | 0.226 | | |
| | Vocal_Surprise | 0.278 | **4.519** | <.001 | 1.060 | 2.701 | | |
| | Vocal_Anger | -0.273 | **-4.466** | <.001 | -4.995 | -1.935 | | |
| | Vocal_Happiness | 0.279 | **4.458** | <.001 | 0.584 | 1.510 | | |
| | Vocal_Fear | 0.018 | 0.305 | 0.761 | -0.496 | 0.677 | | |
| | Vocal_Sadness | 0.025 | 0.405 | 0.686 | -0.406 | 0.616 | | |
| | Vocal_Neutral | -0.056 | -0.928 | 0.355 | -0.447 | 0.161 | | |

**Bold:** $p < .001$.

## 5. DISCUSSION

With the rise in popularity of MOOCs, asynchronous video learning formats, and hybrid models blending face-to-face or synchronous live sessions in adult learning and higher education, the major criticism has been their lack of interactivity and emotional depth (Alemayehu & Chen, 2023). This gap often results in diminished student affective engagement. Many scholars have noted that the vocal emotive expressions of teachers in MOOCs or instructional videos can enhance student affective engagement and should thus be a focus of future research (e.g., Chiu et al., 2022; Ma et al., 2023; Zeng & Luo, 2023).

This study explores and examines whether verbal and nonverbal vocal emotive expressions by teachers can significantly enhance students' affective engagement in asynchronous video learning. Through hypothesis testing using hierarchical linear regression, this study obtains several key findings and insights.

### 5.1 Teachers' verbal vocal emotive expressions do not influence student affective engagement

Contrary to the initial assumption in H1, this study reveals that teachers' verbal emotive expressions (i.e., choice of words) in videos have a negligible effect on student affective engagement. This finding aligns with dual coding theory, which posits that verbal cues engage cognitive processes, whereas nonverbal cues trigger emotional reactions (Paivio, 1971). Thus, spoken words barely influence affective engagement in videos with both cue types, highlighting the profound impact of nonverbal vocal expressions (Scherer, 1986). Hence, the key to increasing the level of student affective engagement in asynchronous video learning lies in leveraging nonverbal vocal expressions rather than leveraging mere verbal content, marking a significant pivot in instructional video design.

### 5.2 Teachers' nonverbal vocal emotive expressions influence student affective engagement

In support of H2 and H3, our study reveals that teachers' nonverbal vocal expressions of happiness and surprise, while mitigating displays of anger, significantly increase student affective engagement in instructional videos. These findings align with theories from paralanguage (Trager, 1958), self-determination (Ryan & Deci, 2000), and voice and emotion (Juslin & Laukka, 2003), emphasizing that teachers can influence students' emotions through nonverbal vocal cues in asynchronous video learning. Our study reveals that emotional contagion is driven more by emotional arousal than by valence alone. Therefore, nonverbal vocal expressions with distinct valence and greater arousal, such as happiness, anger, and surprise, are most effective in influencing student affective engagement, with surprise having the greatest impact, aligning with the circumplex model of affect (Russell, 1980).

5.3 Teachers' nonverbal vocal emotive expressions influence student affective engagement more than do verbal vocal emotive expressions

As in H4, linear regressions indicate that these three types of nonverbal vocal emotive expressions have significantly greater explanatory power for student affective engagement than do verbal emotive expressions, according to dual coding theory (Paivio, 1971) and Mehrabian's (1971) rule. Aside from textual and visual content, students are less likely to focus on the teacher's spoken words but are more influenced by the teacher's nonverbal vocal emotions in this learning mode. The importance of social presence elements in the CoI model (Borup et al., 2012) supports this finding, suggesting that effective nonverbal cues enhance the sense of connection in digital learning. Additionally, the circumplex model of affect (Russell, 1980) highlights the role of emotional arousal in engagement. Teachers should vary their tone to convey enthusiasm through vocal expressions of surprise and happiness, modulate their pitch to emphasize key points, and use varied rhythms to maintain student interest. Reducing the degree of vocal anger is crucial to fostering an affectively engaged learning environment.

5.4 Research limitations and future opportunities

Despite offering numerous new findings and insights, this study acknowledges several research limitations that must be addressed. First, our sample leans toward female Chinese speakers and social science teachers. Future work could diversify the sample and subjects more broadly to improve generalizability.

Second, affective engagement is measured via self-reports, which are prone to bias. Future research could use computer-based emotional recognition tools for more objective assessments (see Mutawa & Sruthi, 2023).

Finally, we analyze the emotive valence and arousal in teachers' speech using the Chinese EmoBank. Future research could implement more sophisticated sentiment analysis techniques, preferring segment analysis to bags of words (Su et al., 2024; Suresh et al., 2024; Zhou & Gao, 2023).

# 6. CONCLUSIONS

While acknowledging its limitations, this study contributes to the field of Human-Computer Interaction, highlighting the critical role of nonverbal vocal emotive expressions—such as happiness, surprise, and anger—in enhancing user affective engagement in asynchronous video environments. The findings provide theoretical insights and practical applications by shifting the focus from verbal content to emotional cues and emphasizing the importance of emotional arousal over valence. The study integrates established theories and frameworks to advance the understanding of how vocal expressions influence social presence and emotional engagement in one-way video-based instruction and communication mediums. Practically, it offers actionable recommendations for leveraging technology and designing interactive systems that optimize vocal emotive expressions, enabling instructors to create more engaging and emotionally resonant learning experiences. These insights address critical gaps in asynchronous learning, paving the way for more emotionally connected and interactive online education.

## 6.1 Research implications

Previous research has greatly enhanced our understanding of affective engagement in MOOCs, focusing on aspects such as student profiles (Shi et al., 2024), feedback mechanisms (Vilkova & Shcheglova, 2021), motivational strategies (Karabatak & Polat, 2020), and accessibility (Mohd Ashril et al., 2024).

This study extends the work of Suen and Hung (2024), who investigated the impact of teachers' facial and paraverbal expressions on students' affective engagement in MOOCs. Expanding their scope, our research focuses specifically on the influence of teachers' verbal and nonverbal vocal emotive expressions in asynchronous video learning environments. By examining verbal elements such as valence (emotional tone) and arousal (intensity) alongside nonverbal vocal emotions like happiness, surprise, and anger, this study aims to provide a more detailed understanding of how these vocal expressions affect students' emotional engagement. Additionally, we address existing research gaps and align with future directions suggested by many scholars (e.g. Chiu et al., 2022; Ma et al., 2023; Zeng & Luo, 2023), offering new insights into optimizing teacher-student interaction and engagement in asynchronous instructional settings.

In addition to employing emotional response theory (Mottet et al., 2006) and emotion contagion theory (Hatfield et al., 1993) to highlight the importance of teacher vocal emotive expressions in classrooms and synchronous online learning environments, this study applies dual coding theory (Paivio, 1971) and Mehrabian's rule (1971) to emphasize the significant impact of teachers' nonverbal vocal emotive expressions by valence and arousal dimensions in prerecorded educational videos. This application provides a novel theoretical basis for examining the relative importance of vocal emotional cues in asynchronous video learning.

To further broaden our theoretical perspective, we incorporate insights from paralanguage (Trager, 1958), self-determination theory (Ryan & Deci, 2000), and voice and emotion theory (Juslin & Laukka, 2003). This integrated approach allows us to identify the key nonverbal vocal emotions—surprise, happiness, and anger—significantly influencing student affective engagement in this context.

Moreover, our multifaceted theoretical framework integrates the social presence elements of the CoI framework (Borup et al., 2012) and the circumplex model of affect (Russell, 1980) to enhance the understanding of how teacher vocal expressions impact student affective engagement in this setting. This contribution extends the literature on the affective dimensions of prerecorded video instruction to encompass the unique characteristics of teacher emotional expressions.

## 6.2 Practical implications

Our study suggests a pivotal shift in focus toward optimizing teachers' vocal emotive expressions in videos to enhance student affective engagement. To begin, video designers must prioritize adaptive training programs. These programs should equip instructors and social media influencers with the skills to modulate their voices effectively. The focus should be on conveying emotions like happiness and surprise while minimizing expressions of anger. This training could involve workshops or coaching sessions focused on vocal techniques that make video content more engaging and emotionally resonant with students and audiences.

Second, video content development should more heavily consider the emotional impact of these nonverbal cues. This might include emphasizing these aspects through postproduction techniques or generative artificial intelligence generated speech and providing guidelines for teachers and speakers—or even replacing their voices—to present videos that maximize positive emotional contagion.

Finally, video content developers might explore new features that enhance the emotional expressiveness of instructors or influencers on video platforms. For example, AI-driven feedback tools could analyze a video's vocal qualities and suggest improvements to teachers or speakers, helping them adjust their delivery for greater emotional impact based on their expressive styles (Zhou & Gao, 2023).

## Acknowledgments

This study was supported by the National Science and Technology Council, Taiwan, under grant NSTC 113-2622-H-003 -003 -, NSTC 112-2410-H-003-102-MY2, and NSTC 111-2410-H-019-006-MY3. This work was financially/partially supported by the Advanced Institute of Manufacturing with High-tech Innovations (AIM-HI) from the Featured Areas Research Center Program within the framework of the Higher Education Sprout Project by the Ministry of Education (MOE) in Taiwan.

We also extend our gratitude to Chin-Chia Yeh, Chia-Fan Chu, Jing-Rui Gu, Kuo-En Hung, Yan-Ming Huang, Yi-Chen Lin, Yu-Cheng Tseng, and Yung-Sian Fang from National Taiwan Normal University for their assistance in data collection, data cleaning, and pilot testing for this study.

## References

Alemayehu, L., & Chen, H.-L. (2023). Learner and instructor-related challenges for learners' engagement in MOOCs: A review of 2014–2020 publications in selected SSCI indexed journals. *Interactive Learning Environments*, *31*(5), 3172-3194. https://doi.org/10.1080/10494820.2021.1920430

Allison, T. H., Warnick, B. J., Davis, B. C., & Cardon, M. S. (2022). Can you hear me now? Engendering passion and preparedness perceptions with vocal expressions in crowdfunding pitches. *Journal of Business Venturing*, *37*, 106193. https://doi.org/10.1016/j.jbusvent.2022.106193

Akiha, K., Brigham, E., Couch, B. A., Lewin, J., Stains, M., Stetzer, M. R., Vinson, E. L., & Smith, M. K. (2018). What types of instructional shifts do students experience? Investigating active learning in science, technology, engineering, and math classes across key transition points from middle school to the university level. *Frontiers in Education*, *2*, 68. https://doi.org/10.3389/feduc.2017.00068

Bao, W., Li, Y., Gu, M., Yang, M., Li, H., Chao, L., & Tao, J. (2014). Building a chinese natural emotional audio-visual database. In 2014 12th International Conference on Signal Processing (ICSP) (pp. 583-587). IEEE. https://doi.org/10.1109/ICOSP.2014.7015071

Beukeboom, C. J., & Semin, G. R. (2006). How mood turns on language. *Journal of Experimental Social Psychology*, *42*, 553–566. https://doi.org/10.1016/j.jesp.2005.09.005

Borup, J., West, R. E., & Graham, C. R. (2012). Improving online social presence through asynchronous video. *Internet and Higher Education*, *15*, 195–203. https://doi.org/ 10.1016/j.iheduc.2011.11.001

Cavanagh, M., Bower, M., Moloney, R., & Sweller, N. (2014). The effect over time of a video-based reflection system on preservice teachers' oral presentations. *Australian Journal of Teacher Education*, *39*(6), Article 1. https://doi.org/10.14221/ajte.2014v39n6.3

Chew, Y. W. (2022). Performing presence with the Teaching-body via videoconferencing: A postdigital study of the Teacher's face and Voice. *Postdigital Science and Education*, *4*, 394–42. https://doi.org/10.1007/s42438-022-00288-2

Chiu, T. K. F. (2022). Applying the self-determination theory (SDT) to explain student engagement in online learning during the COVID-19 pandemic. *Journal of Research on Technology in Education*, 54, 14–30. https://doi.org/10.1080/15391523.2021.1891998

Daher, W., Sabbah, K., & Abuzant, M. (2021). Affective engagement of higher education students in an online course. *Emerging Science Journal*, *5*, 545–558. https://doi.org/10.28991/esj-2021-01296

Dai, H. M., Teo, T., & Rappa, N. A. (2020). Understanding continuance intention among MOOC participants: The role of habit and MOOC performance. *Computers in Human Behavior*, *112*, 106455. https://doi.org/10.1016/j.chb.2020.106455.

Dixson, M. D. (2015). Measuring student engagement in the online course: The Online Student Engagement scale (OSE). *Online Learning*, *19*(4), 165. https://doi.org/10.24059/olj.v19i4.561

Dixson, M. D. (2010). Creating effective student engagement in online courses: What do students find engaging? *Journal of the Scholarship of Teaching and Learning*, *10*, 1–13. https://scholarworks.iu.edu/journals/index.php/josotl/article/view/1744

El Ayadi, M., Kamel, M. S., & Karray, F. (2011). Survey on speech emotion recognition: Features, classification schemes, and databases. *Pattern Recognition*, *44*, 572–587. https://doi.org/10.1016/j.patcog.2010.09.020

Fabriz, S., Mendzheritskaya, J., & Stehle, S. (2021). Impact of synchronous and asynchronous settings of online teaching and learning in higher education on students' learning experience during COVID-19. *Frontiers in Psychology*, *12*, 733554. https://doi.org/ 10.3389/fpsyg.2021.733554

Fanselow, M. S. (2018). Emotion, motivation and function. *Current Opinion in Behavioral Sciences*, *19*, 105–109. https://doi.org/10.1016/j.cobeha.2017.12.013

Farrell, O., & Brunton, J. (2020). A balancing act: A window into online student engagement experiences. *International Journal of Educational Technology in Higher Education*, *17*(25). https://doi.org/10.1186/s41239-020-00199-x

Garcia, M., & Yousef, A. (2023). Cognitive and affective effects of teachers' annotations and talking heads on asynchronous video lectures in a web development course. *Research and Practice in Technology Enhanced Learning*, *18*(20), 1–23. https://doi.org/10.58459/rptel.2023.18020.

Gunes, H., Schuller, B., Pantic, M., & Cowie, R. (2011). Emotion representation, analysis and synthesis in continuous space: A survey. *Proceedings of the 2011 IEEE International Conference on Automatic Face & Gesture Recognition (FG)*, 827-834. Santa Barbara, CA, USA. https://doi.org/10.1109/FG.2011.5771357

Hatfield, E., Cacioppo, J. T., & Rapson, R. L. (1993). Emotional contagion. *Current directions in psychological science*, *2*, 96–100. https://doi.org/10.1111/1467-8721.ep10770953

Heikinheimo, H. (2023). Analyzing open AI's whisper ASR accuracy: Word error rates across languages and model sizes. https://www.speechly.com/blog/analyzing-open-ais-whisper-asr-models-word-error-rates-across-languages

Henderson, M. L., & Schroeder, N. L. (2021). A Systematic review of instructor presence in instructional videos: Effects on learning and affect. *Computers and Education Open*, *2*, 100059. https://doi.org/10.1016/j.caeo.2021.100059.

Henrie, C. R., Halverson, L. R., & Graham, C. R. (2015). Measuring student engagement in technology-mediated learning: A review. *Computers & Education*, *90*, 36–53. https://doi.org/10.1016/j.compedu.2015.09.005

Horan, S. M., Martin, M. M., & Weber, K. (2012). Understanding emotional response theory: The role of instructor power and justice messages. *Communication Quarterly*, *60*, 210–233. https://doi.org/10.1080/01463373.2012.669323

Hsu, J. H., Su, M. H., Wu, C. H., & Chen, Y. H. (2021). Speech emotion recognition considering nonverbal vocalization in affective conversations. *IEEE/ACM Transactions on Audio, Speech, and Language Processing*, *29*, 1675–1686. https://doi.org/10.1109/TASLP.2021.3076364

Immordino-Yang, M. H., & Damasio, A. (2007). We feel, therefore we learn: The relevance of affective and social neuroscience to education. *Mind, Brain, and Education*, *1*(1), 3-10. https://doi.org/10.1111/j.1751-228X.2007.00004.x

Iskrenovic-Momcilovic, O. (2018). Learning a programming language. *International Journal of Electrical Engineering Education*, *55*, 324–333. https://doi.org/10.1177/0020720918773975

Jieba-tw (2023). Jieba Chinese word segmentation Taiwan traditional Chinese version. https://github.com/APCLab/jieba-tw

Juslin, P. N., & Laukka, P. (2003). Communication of emotion in vocal expression and music performance: Different channels, same code? *Psychological Bulletin*, *129*, 770–814. https://doi.org/10.1037/0033-2909.129.5.770

Karabatak, S., & Polat, H. (2020). The effects of the flipped classroom model designed according to the ARCS motivation strategies on the students' motivation and academic achievement levels. *Education and Information Technologies*, *25*, 1475–1495. https://doi.org/10.1007/s10639-019-09985-1

Kraus, M. W. (2017). Voice-only communication enhances empathic accuracy. *American Psychologist*, *72*, 644–654. https://doi.org/10.1037/amp0000147

Laukka, P. (2017). Vocal communication of emotion. In: V. Zeigler-Hill & T. Shackelford (eds) *Encyclopedia of Personality and Individual Differences*. Springer, Cham. Retrieved from https://doi.org/10.1007/978-3-319-28099-8_562-1

Lee, L.-H., Li, J.-H., & Yu, L.-C. (2022). Chinese EmoBank: Building valence-arousal resources for dimensional sentiment analysis. *ACM Transactions on Asian and Low-Resource Language Information Processing*, *21*(4), 1–18. https://doi.org/10.1145/3489141

Lee, S., & Potter, R. F. (2020). The impact of emotional words on listeners' emotional and cognitive responses in the context of advertisements. *Communication Research*, *47*, 1155–1180. https://doi.org/10.1177/0093650218765523

Leung, Y., Oates, J., & Chan, S. (2018). Voice, articulation, and prosody contribute to listener perceptions of speaker gender: A systematic review and meta-analysis. *Journal of speech, language, and hearing research*, *61*(2), 266–297. https://doi.org/10.1044/2017_JSLHR-S-17-0067.

Liebenthal, E., Silbersweig, D. A., & Stern, E. (2016). The language, tone and prosody of emotions: neural substrates and dynamics of spoken-word emotion perception. *Frontiers in Neuroscience*, *10*. https://doi.org/10.3389/fnins.2016.00506

Lin, L. C., Hung, I. C., Kinshuk, K., & Chen, N. S. (2019). The impact of student engagement on learning outcomes in a cyber-flipped course. *Education Technology Research Development*, 67, 1573–1591. https://doi.org/10.1007/s11423-019-09698-9

Liu, X.-Y., Chi, N.-W., & Gremler, D. D. (2019). Emotion cycles in services: Emotional contagion and emotional labor effects. *Journal of Service Research*, *22*, 285–300. https://doi.org/10.1177/1094670519835309

Ma, N., Zhang, Y.-L., Liu, C.-P., & Du, L. (2023). The comparison of two automated feedback approaches based on automated analysis of the online asynchronous interaction: A case of massive online teacher training. *Interactive Learning Environments*. https://doi.org/10.1080/10494820.2023.2191252

Martin, F., & Bolliger, D. U. (2018). Engagement matters: Student perceptions on the importance of engagement strategies in the online learning environment. *Online Learning*, *22*, 205–222. https://doi.org/10.24059/olj.v22i1.1092

Maximize Market Research. (2023). *MOOC market - global industry analysis and forecast (2023-2029*). https://www.maximizemarketresearch.com/market-report/mooc-market/120455/

Mehrabian, A. (1971). Silent messages. Wadsworth.

Mershad, K., & Said, B. (2022). DIAMOND: A tool for monitoring the participation of students in online lectures. *Education and Information Technologies*, *12*, 1–31. https://doi.org/10.1007/s10639-021-10801-y

Mohd Ashril, N. A. N., Chee, K. N., Yahaya, N., & Abdul Razak, R. (2024). Barriers, strategies and accessibility: Enhancing engagement and retention of learners with disabilities in MOOCs – A systematic literature review (SLR). *International Journal of Human–Computer Interaction*, 1–12. https://doi.org/10.1080/10447318.2024.2414892

Mottet, T. P., Richmond, V. P., & McCroskey, J. C. (2006). *Handbook of instructional communication: Rhetorical and relational perspectives*. Allyn & Bacon. https://doi.org/10.4324/9781315189864

Mutawa, A. M., & Sruthi, S. (2023). Enhancing human–computer interaction in online education: A machine learning approach to predicting student emotion and satisfaction. *International Journal of Human–Computer Interaction*, *40*(24), 8827–8843. https://doi.org/10.1080/10447318.2023.2291611

OpenAI. (2022). Introducing Whisper. Retrieved from https://openai.com/research/whisper

Paivio, A. (1971). *Imagery and verbal processes*. Holt, Rinehart, and Winston.

Perry, M., Azevedo, R. F. L., Henricks, G., Crues, R. W., & Bhat, S. (2024). Learning from online instructional videos considering video presentation modes, technological comfort, and student characteristics. *International Journal of Human–Computer Interaction*, 1–18. https://doi.org/10.1080/10447318.2024.232891

Richardson, J. C., & Lowenthal, P. (2017). Instructor social presence: Learners' needs and a neglected component of the community of inquiry framework. In A. Whiteside, A. Garrett Dikkers, & K. Swan, (Eds.), Social presence in online learning: Multiple perspectives on practice and research (pp. 86-98). Stylus.

Russell, J. A. (2003). Core affect and the psychological construction of emotion. *Psychological Review*, *110*, 145–172. https://doi.org/10.1037/0033-295X.110.1.145

Russell, J. A. (1980). A circumplex model of affect. *Journal of Personality and Social Psychology*, *39*(6), 1161–1178. https://doi.org/10.1037/h0077714

Ryan, R. M., & Deci, E. L. (2000). Self-determination theory and the facilitation of intrinsic motivation, social development, and well-being. *American Psychologist*, *55*(1), 68–78. https://doi.org/10.1037/0003-066X.55.1.68

Sharanyaa, S., Mercy, T. J., & V.G, S. (2023). Emotion recognition using speech processing. In *2023 3rd International Conference on Intelligent Technologies (CONIT)* (pp. 1-5). https://doi.org/10.1109/CONIT59222.2023.10205935

Sauter, D. A., Eisner, F., Calder, A. J., & Scott, S. K. (2010). Perceptual cues in nonverbal vocal expressions of emotion. *The Quarterly Journal of Experimental Psychology*, *63*, 2251–2272. https://doi.org/10.1080/17470211003721642

Scherer, K. R. (1986). Vocal affect expression: A review and a model for future research. *Psychological Bulletin*, *99*(2), 143–165. https://doi.org/10.1037/0033-2909.99.2.143

Schuller, B., Batliner, A., Steidl, S., Seppi, D., Zhang, Y., Weninger, F., ... & Sethu, V. (2018). The interspeech 2018 computational paralinguistics challenge: Atypical & self-assessed affect, crying & heart beats. *In Proceedings of Interspeech 2018* (pp. 122-126). https://doi.org/10.21437/Interspeech.2018-1548

Shi, H., Zhou, Y., Dennen, V. P., et al. (2024). From unsuccessful to successful learning: Profiling behavior patterns and student clusters in Massive Open Online Courses. *Education and Information Technologies*, *29*, 5509–5540. https://doi.org/10.1007/s10639-023-12010-1

Simon-Thomas, E. R., Keltner, D. J., Sauter, D., Sinicropi-Yao, L., & Abramson, A. (2009). The voice conveys specific emotions: Evidence from vocal burst displays. *Emotion*, *9*, 838 – 846. http://dx.doi.org/10.1037/a0017810

Stevenson, R. A., Mikels, J. A., & James, T. W. (2007). Characterization of the affective norms for English words by discrete emotional categories. *Behavior Research Methods*, *39*, 1020–1024. https://doi.org/10.3758/bf03192999

Su, Y. S., Hu, Y. C., Wu, Y. C., & Lo, C. T. (2024). Evaluating the impact of pumping on groundwater level prediction in the chuoshui river alluvial fan using artificial intelligence techniques. *International Journal of Interactive Multimedia and Artificial Intelligence*, 8(7), 28-37. https://doi.org/10.9781/ijimai.2024.04.002

Su Y.S., Lin Y.D., & Liu T.Q. (2022). Applying machine learning technologies to explore students' learning features and performance prediction. *Frontiers in Neuroscience*. 16, 1018005. https://doi.org/10.3389/fnins.2022.1018005_Su

Su Y.S., Ding T.J., & Chen M.Y. (2021a). Deep Learning Methods in Internet of Medical Things for Valvular Heart Disease Screening System. *IEEE Internet of Things Journal*, 8(23), 16921-16932. https://doi.org/10.1109/JIOT.2021.3053420

Su Y.S., Suen H.Y., & Hung K.E. (2021b). Predicting behavioral competencies automatically from facial expressions in real-time video-recorded interviews. *Journal of Real-Time Image Processing*, 18(4), 1011-1021. https://doi.org/10.1007/s11554-021-01071-5

Su, Y., Zhang, K., Wang, J., Zhou, D., & Madani, K. (2020). Performance analysis of multiple aggregated acoustic features for environment sound classification. *Applied Acoustics*, *158*, 107050. https://doi.org/10.1016/j.apacoust.2019.107050.

Suen, H. Y., & Hung, K. E. (2024). Enhancing learner affective engagement: The impact of instructor emotional expressions and vocal charisma in asynchronous video-based online learning. *Education and Information Technologies*. https://doi.org/10.1007/s10639-024-12956-w

Suresh, T., Sengupta, A., Akhtar, M. S., & Chakraborty, T. (2024). A comprehensive understanding of code-mixed language semantics using hierarchical transformer. *IEEE Transactions on Computational Social Systems*, *11*(3), 4139–4148. https://doi.org/10.1109/TCSS.2024.3360378

Trager, G. L. (1958). Paralanguage: A first approximation. *Studies in Linguistics*, *13*, 1-12.

Tursunov, A., Kwon, S., & Pang, H. S. (2019). Discriminating emotions in the valence dimension from speech using timbre features. *Applied Sciences*, 9(12), 2470. https://doi.org/10.3390/app9122470

Walker, K. A., & Koralesky, K. E. (2021). Student and instructor perceptions of engagement after the rapid online transition of teaching due to COVID-19. *Natural Sciences Education*, 50, e20038. https://doi.org/10.1002/nse2.20038

Wang, Y. (2022). To be expressive or not: the role of teachers' emotions in students' learning. *Frontiers in Psychology*, *12*, 737310. https://doi.org/10.3389/fpsyg.2021.737310

Wang, M., Chen, Z., Shi, Y., Wang, Z., & Xiang, C. (2022). Instructors' expressive nonverbal behavior hinders learning when learners' prior knowledge is low. *Frontiers in Psychology*, 13. https://doi.org/10.3389/fpsyg.2022.810451.

Weiss, H. M., & Cropanzano, R. (1996). Affective events theory: A theoretical discussion of the structure, causes and consequences of affective experiences at work. In B. M. Staw & L. L. Cummings (Eds.), Research in organizational behavior: An annual series of analytical essays and critical reviews, Vol. 18 (pp. 1-74). Elsevier Science/JAI Press.

Ye, B., Yuan, X., Peng, G., & Zeng, W. (2022). A Novel Speech Emotion Model Based on CNN and LSTM Networks. In *2022 6th Asian Conference on Artificial Intelligence Technology (ACAIT)*, 1-4. https://doi.org/10.1109/ACAIT56212.2022.10137926

Vilkova, K., & Shcheglova, I. (2021). Deconstructing self-regulated learning in MOOCs: In search of help-seeking mechanisms. *Education and Information Technologies*, *26*, 17–33. https://doi.org/10.1007/s10639-020-10244-x

Vo, H., & Ho, H. (2024). Online learning environment and student engagement: The mediating role of expectancy and task value beliefs. *Australian Educational Researcher*. https://doi.org/10.1007/s13384-024-00689-1

Yuan, M., Zeng, J., Wang, A., & Shang, J. (2021). Would it be better if instructors technically adjust their image or voice in online courses? Impact of the way of instructor presence on online learning. *Frontiers in Psychology*, *12*, 746857. https://doi.org/10.3389/fpsyg.2021.746857

Zeng, H., & Luo, J. (2023). Effectiveness of synchronous and asynchronous online learning: A meta-analysis. *Interactive Learning Environments*, 1–17. https://doi.org/10.1080/10494820.2023.2197953

Zhao, H. & Khan, A. (2022). The students' flow experience with the continuous intention of using online English platforms. *Frontiers in Psychology*, *12*, 807084. https;//doi.org/10.3389/fpsyg.2021.807084

Zhou, W., & Gao, B. (2023). Construction and application of English-Chinese multimodal emotional corpus based on artificial intelligence. *International Journal of Human–Computer Interaction*, 1–12. https://doi.org/10.1080/10447318.2023.2169526

## About the authors

**Hung-Yue Suen** is a Professor in the Department of Technology Application and Human Resource Development at National Taiwan Normal University. His research interests center on affective computing, human-AI interaction (HAI), human resource technology, and AI-enabled talent assessment.

**Yu-Sheng Su** is an Associate Professor at the Department of Computer Science and Information Engineering, National Chung Cheng University, Taiwan. His research focus includes cloud computing, big data analytics, intelligent systems, and the metaverse.